\documentclass[twocolumn,showpacs,preprintnumbers,amsmath,amssymb]{revtex4}
\usepackage{graphicx}% Include figure files
\usepackage{dcolumn}% Align table columns on decimal point
\usepackage{bm,color}% bold math
\def\jpsi{{J/\psi}}

\def\be{\begin{equation}}
\def\ee{\end{equation}}
\def\bea{\begin{eqnarray}}
\def\eea{\end{eqnarray}}
\def\NO{\nonumber}
\def\gev{\mathrm{~GeV}}
\def\kev{\mathrm{~KeV}}

\def\dfrac{\displaystyle\frac}
\def\a{\alpha}

\def\s{\sigma}

\begin{document}

%\preprint{APS/123-QED}

\title{Testing the QCD fragmentation mechanism on heavy quarkonium production at LHC}%  \\

\author{Bin Gong$^{1,2,3}$, Rong Li$^{2,3}$ and Jian-Xiong Wang$^{2,3}$}%
\affiliation{
Institute of Theoretical Physics, CAS, P.O. Box 2735, Beijing, 100190, China. \\
Institute of High Energy Physics, CAS, P.O. Box 918(4), Beijing, 100049, China. \\
%}
%\affiliation{
Theoretical Physics Center for Science Facilities, CAS, Beijing, 100049, China.
}%
\date{\today}% It is always \today, today,

\begin{abstract}{\label{abstract}}
We calculate the fragmentation function for charm quark into $\jpsi$ at the QCD next-to-leading-order (NLO)
and find that the produced $\jpsi$ is of larger momentum fraction than it is at the leading-order.
Based on the fragmentation function and partonic processes calculated at the NLO, the transverse momentum
distribution on $J/\psi$ hadroproduction associated with a charm c (or $\bar{c}$) jet are predicted.
We find that the distribution is enhanced by a factor of 2.0$\sim$3.3 at the NLO as $p_t$ increased from 10 GeV to 100 GeV and it is measurable at the LHC with charm tagger. The measurement at the LHC will supply a first chance to directly
test the QCD fragmentation mechanism on heavy quarkonium production where the fragmentation function is calculable
in perturbative QCD.  It is also applied to $\jpsi$ ($\Upsilon$) production in the decay of $Z^0$ (top quark).
\end{abstract}

                             % Classification Scheme.in \cite{Bodwin:2002kk}
\maketitle
{\label{introduction}}

   Quantum Chromodynamics (QCD) is a successful theory to describe strong
interaction, but its fundamental ingredients, the quarks and gluons, are not observed freely and must hadronize eventually.
The fact makes it impossible to calculate
any processes involving detected hadrons in the final or initial states directly.
According to the QCD factorization theorem (see Ref.~\cite{Collins:1989gx}
and references therein),
in some kinematical regions, the dominant
contribution to the cross section can be decomposed into three parts: the partonic part,
the part of parton fragmentation into the produced hadron and the part of the parton
distributions in the initial hadrons. The partonic part can be calculated perturbatively
because of the asymptotic freedom of QCD, while all the long distance physics of
the hadrons is put into the parton fragmentation functions and the parton distributions.
Therefore, fragmentation functions is one of the most important ingredients to understand
QCD or to make predictions for experimental measurements. It is hard to study fragmentation
functions directly from QCD because of their non-perturbative nature.

For light hadrons, the fragmentation functions are extracted from global data fits.
Recently the transverse-momentum $p_T$ distribution of inclusive light-charged-particle production
measured by the CDF shows significantly exceed on the theoretical prediction based on these
fragmentation functions when $p_t>80\gev$ \cite{Albino:2010em}.
It potentially challenges our understanding of QCD factorization theorem.
However, for heavy quarkonium, the non-relativistic QCD (NRQCD) factorization formalism
\cite{Bodwin:1994jh} can be used to factorize the
fragmentation functions for quarkonium into NRQCD matrix elements and short-distance factors,
which are calculable in perturbation theory  and have been studied
in many works \cite{Braaten:1993rw, Braaten:1993mp,Chen:1993ii}
%{,Ma:1994zt,Ma:1995vi,Qi:2007sf}
at QCD leading-order (LO). There are also QCD next-to-leading order (NLO) studies on
the color-octet $^3S_1$ gluon fragmentation function for heavy quarkonium~\cite{Ma:1995ci}, %{,Beneke:1995yb,Braaten:2000pc}
and the study on relativistic corrections for
the fragmentation functions~\cite{Sang:2009zz}.
The QCD factorization on heavy quarkonium production is investigated by many authors~\cite{Nayak:2005rw}.%{,Nayak:2005rt,Nayak:2007mb,Bodwin:2008nf,Bodwin:2010fi}
Therefore, we have more prediction
power to test QCD fragmentation mechanism on heavy quarkonium production.
But until now, there is no available experimental measurement to test it.

In recent years, there is a huge data collection at colliders. Based on that, many $\jpsi$
production processes were observed~\cite{Abe:2001za}
%{,Abe:2002rb,Aubert:2005tj}
in the past. Therefore it supplies a very important chance to perform
systematical study on $\jpsi$ production both theoretically and experimentally. The large
discrepancies for exclusive $\jpsi$ productions at the $B$ factories have been studied and
resolved by introducing higher order corrections
~\cite{Braaten:2002fi}.
%{,Bodwin:2002kk,Zhang:2005ch,Gong:2007db,Gong:2008ce}
The discrepancies for inclusive $\jpsi$ production at
the $B$ factories have also been studied~\cite{Zhang:2006ay,Gong:2009ng}
%{,Ma:2008gq,Gong:2009kp}
and the results including NLO QCD corrections can nearly explain the
experimental data. Higher order corrections for $\jpsi$ production at hadron colliders were also
investigated~\cite{Campbell:2007ws}%{,Gong:2008hk,Gong:2008sn,Gong:2008ft,Gong:2010bk,Ma:2010vd,Ma:2010yw,Butenschoen:2010rq}
. Although large improvement have been achieved in theoretical predictions, the experimental data
are still unable to fully understand, especially for the polarization of $\jpsi$.

Higher order contributions have shown their importance and the dominant production mechanism
for heavy quarkonium is fragmentation at large transverse momentum region. To achieve reasonable theoretical
predictions for $\jpsi$ production at large $p_t$, it is important to study higher order contribution
to fragmentation functions for heavy quarkonium states. Moreover, the most important question here is:
is there any chance to directly test the QCD fragmentation mechanism on heavy quarkonium production at the LHC?
In this letter, we calculate the fragmentation functions for charm quark into
$\jpsi$ at QCD NLO.  Based on the fragmentation function and partonic processes calculated at the NLO,
the transverse momentum distribution on $J/\psi$ production associated with a charm c (or $\bar{c}$) jet
are predicted. We find that the distribution is enhanced by a factor of 2.0$\sim$3.3 at the NLO as $p_t$ increased from 10 GeV to 100 GeV
and it is measurable at the LHC with charm tagger~\cite{Menon:2010vm}.

Here we study  $\jpsi$ production in $e^+e^-\rightarrow \jpsi +X$ with a very high c.m. energy $\sqrt{s}$,
which should be dominated by fragmentation mechanism. The differential cross section for $\jpsi$ with momentum $p$ is
\bea
&&d\sigma[e^+e^-\rightarrow\jpsi(p)+X]\\[3mm]
&=&\sum_i\int dzd\sigma[e^+e^-\rightarrow i(p/z)+X,\mu_F]D_{i\rightarrow\jpsi}\left(z,\mu_F\right).\NO
\eea
Where the factorization scale $\mu_F$ is introduced to maintain this factorized form and the dependence on 
the arbitrary scale $\mu_F$ cancels between the two factors. And $i$ could represent all quarks and gluon, but
we only consider $i=c$ (i.e. $e^+e^-\rightarrow J/\psi+c\bar{c}$ channel) in the theoretical calculation for 
both side of Eq.(1) to extract charm quark c to $J/\psi$ fragmentation function.  Here we choose $\mu_F=3m_c$ 
to avoid large logarithms of $\mu_F/m_c$ in the fragmentation function $D_{c(\bar{c})\rightarrow\jpsi}(z,\mu_F)$. 
Hereafter $D(z)$ is used to represent $D(z,3m_c)$. Thus we have
\bea\label{eqn:all_order}
\dfrac{d\sigma_{e^+e^-\rightarrow\jpsi +X}}{d E_\jpsi}
=2\int \dfrac{dE_c}{E_c}\dfrac{d\sigma_{e^+e^-\rightarrow c+X}}{d E_c} D_{c\rightarrow\jpsi}\left(z\right)
\eea
up to NLO in $\a_s$ with $z=E_\jpsi/E_c$. As mentioned in Refs.~\cite{Braaten:1993rw,Braaten:1993mp}, the initial
fragmentation function can be calculated perturbatively as a series in $\a_s(2m_c)$ and extracted from
Eq.~(\ref{eqn:all_order}) order by order.

At leading-order (LO) in $\a_s$, there are $E_c=E_{\bar{c}}=\sqrt{s}/2$ and $z=2E_\jpsi/\sqrt{s}$, and Eq.~(\ref{eqn:all_order}) is simplified into
\bea
\dfrac{d\sigma^{LO}_{e^+e^-\rightarrow\jpsi + X}}{d E_\jpsi}
=\frac{4}{\sqrt{s}}\sigma^{LO}_{e^+e^-\rightarrow c+X} D^{LO}_{c\rightarrow\jpsi}(z),
\eea
Thus the fragmentation function is extracted as:
\bea
D^{LO}_{c\rightarrow\jpsi}(z)=\frac{1}{\s_{c}^*}\dfrac{d\sigma^{LO}_{e^+e^-\rightarrow\jpsi +X}}{d E_\jpsi},
\s_{c}^*\equiv \frac{4\sigma^{LO}_{e^+e^-\rightarrow c+X}}{\sqrt{s}}.
\label{eqn:LO}
\eea
Under the limitation $m_c/\sqrt{s} \rightarrow 0$,  $\s_{c}^*=(64\pi\alpha^2)/(9s^{3/2})$ is obtained,
and by using Eq.(4) in Ref.~\cite{Gong:2009ng} we obtain
\bea
&&D^{LO}_{c\rightarrow\jpsi}(z)=\dfrac{8\a_s(2m_c)^2|R_s(0)|^2}{27\pi m_c^3} \NO\\
&&\qquad\times\dfrac{z(1-z)^2(16-32z+72z^2-32z^3+5z^4)}{(2-z)^6},
\label{eqn:LO2}
\eea
which is exactly the same as the one obtain from $Z_0$ decay in Ref.~\cite{Braaten:1993mp}.

\begin{figure}
\center{
\includegraphics[scale=0.21]{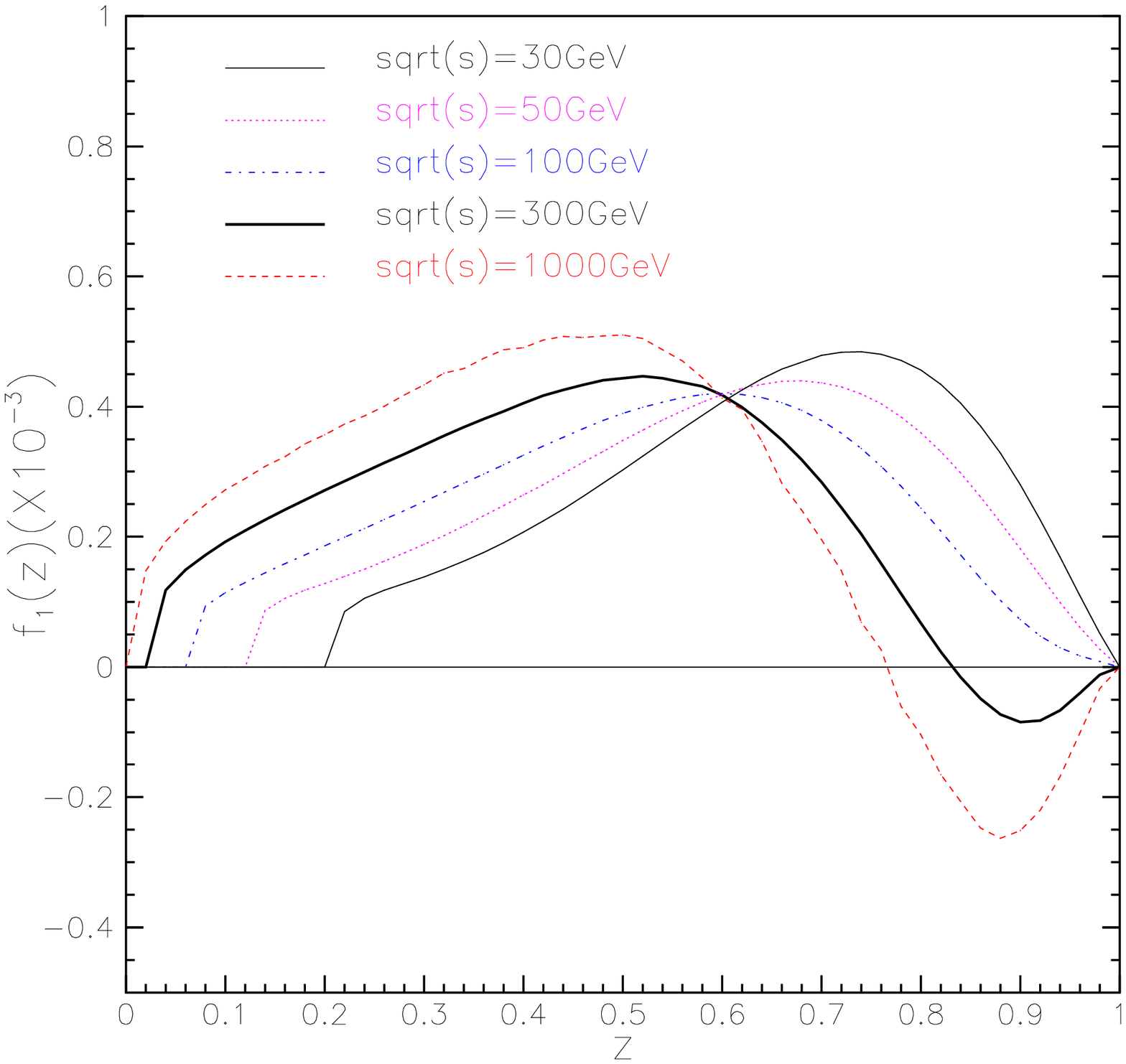}
~~\includegraphics[scale=0.21]{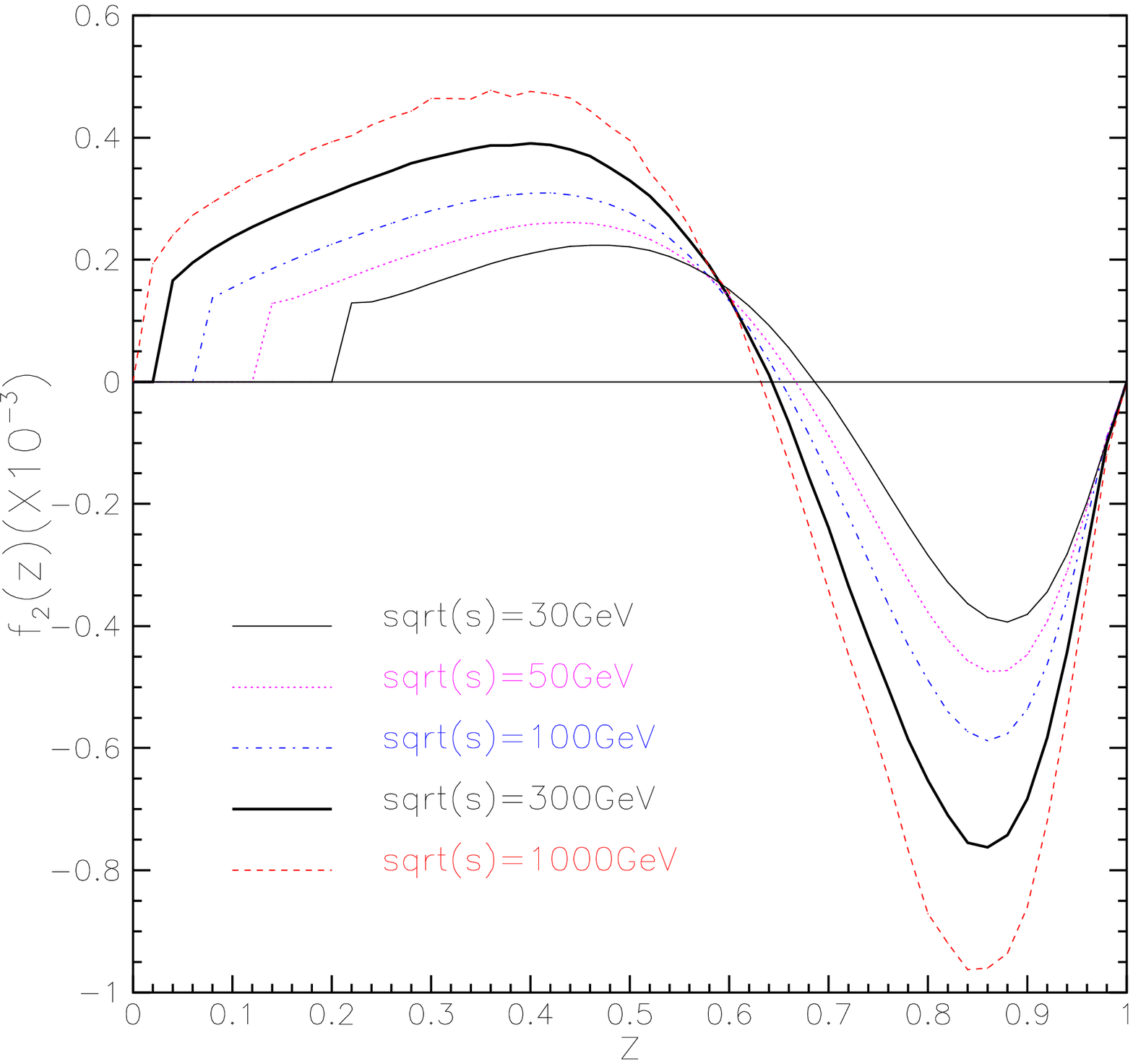}
\caption {\label{fig:f1}Behavior of $f_1(z)$ and $f_2(z)$ with $\mu_F=3m_c$.
}}
\end{figure}

\begin{figure}
\center{
\includegraphics[scale=0.21]{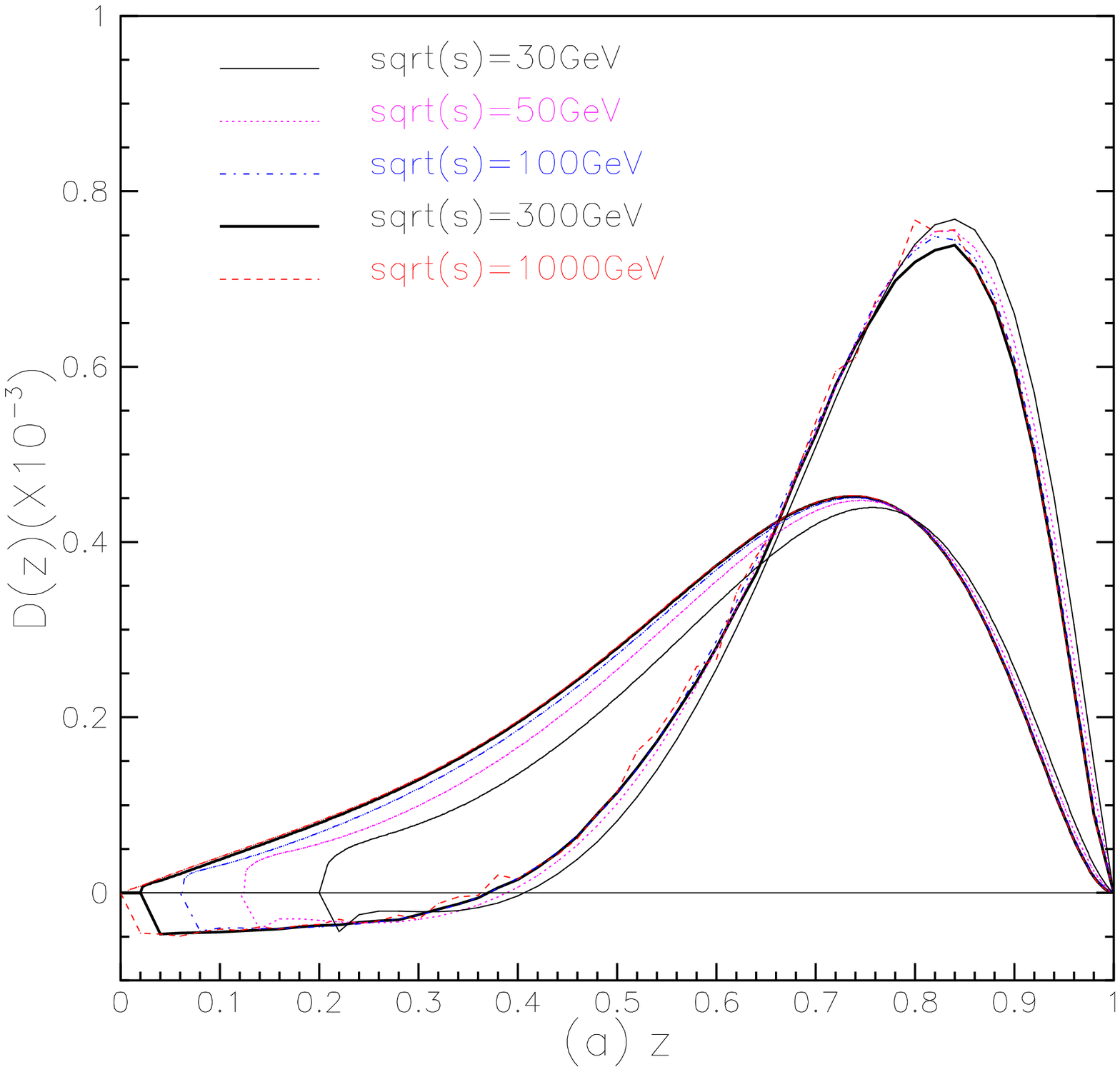}
~~\includegraphics[scale=0.21]{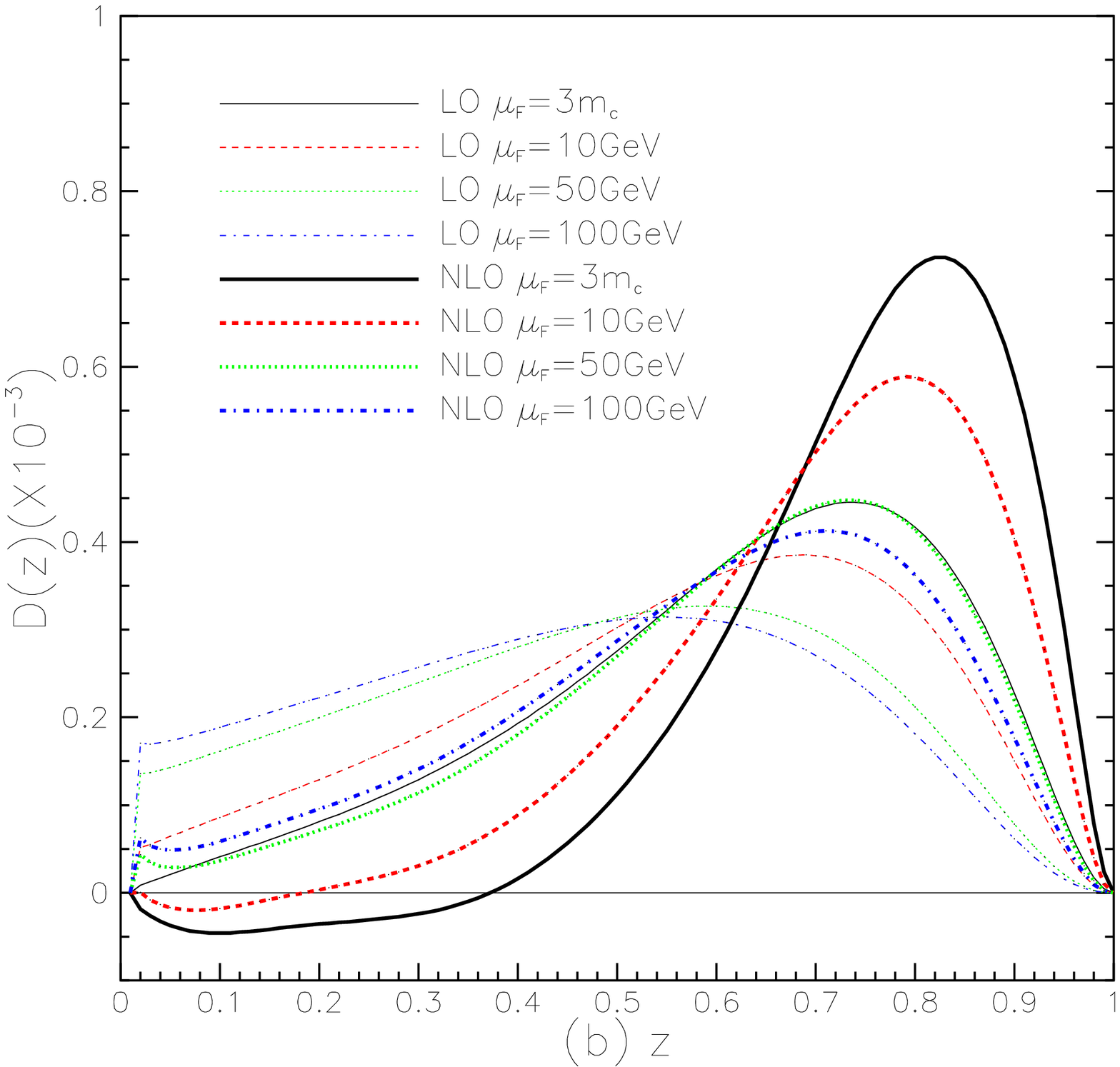}
\caption {\label{fig:frag} The fragmentation functions at LO and NLO (In left figure, $\mu_F$=3$m_c$ is chosen and lower curves are LO ones).
%(the left with $\mu_F$=3$m_c$,  the right with different $\mu_F$).
}}
\end{figure}

In the fix order calculation at NLO in $\a_s$, Eq.~(\ref{eqn:all_order}) becomes
\bea
&\dfrac{d\sigma^{NLO}_{e^+e^-\rightarrow\jpsi +X}}{d E_\jpsi}
=2\int \dfrac{dE_c}{E_c}\dfrac{d\sigma^{LO}_{e^+e^-\rightarrow c+X}}{d E_c} D_{c\rightarrow\jpsi}^{NLO}\left(z\right)\NO\\
&+2\int \dfrac{dE_c}{E_c}\dfrac{d\sigma^{NLO}_{e^+e^-\rightarrow c+X}-d\sigma^{LO}_{e^+e^-\rightarrow c+X}}{d E_c}
D_{c\rightarrow\jpsi}^{LO}\left(z\right).
\label{eqn:NLO}
\eea
Then the NLO fragmentation function is expressed as
\bea
&D_{c\rightarrow\jpsi}^{NLO}(z)=f_1(z)-f_2(z)\\
&f_1(z)=\dfrac{1}{\s_{c}^*}\dfrac{d\sigma^{NLO}_{e^+e^-\rightarrow\jpsi +X}}{d E_\jpsi}\NO\\
&f_2(z)=\dfrac{2}{\s_{c}^*}\int \dfrac{dE_c}{E_c}\dfrac{d(\sigma^{NLO}-\sigma^{LO})_{e^+e^-\rightarrow c+X}}{d E_c}
D_{c\rightarrow\jpsi}^{LO}\left(z\right)\NO
\eea
$f_1$ is the energy distribution of $\jpsi c\bar{c}$ production at NLO, which have been achieved in our previous work.
And $f_2$ needs the LO fragmentation function in Eq.~(\ref{eqn:LO2}).
It is hard to obtain an analytic result here and we have to do it numerically.
We choose $m_c=1.5\gev$, $\a_s(2m_c)=0.26$ and $|R_s(0)|^2=0.944\gev$, and the renormalization scale
$\mu_R=2m_c$. The behaviors of $f_1(z)$ and $f_2(z)$ shown in Fig.~\ref{fig:f1}
strongly depend on $\sqrt{s}$. In Fig.~\ref{fig:frag}a,
the fragmentation functions extracted numerically at the LO and NLO
show very good limitation as $\sqrt{s}$ increasing from $30\gev$ to $1000\gev$ and
even the difference between the $\sqrt{s}=30\gev$ and $\sqrt{s}=1000\gev$ is quite small.
It means that the fragmentation mechanism can describe the theoretical result even when the c.m. energy is as low as $30\gev$.
For the NLO results, there exists unphysical range of negative possibility arising
from fix-order perturbative calculation. The result with $\sqrt{s}=1000\gev$ is a bit unstable 
even in our quadruple precision FORTRAN calculation because 
of large numerical cancellation. Therefore we choose the result at $\sqrt{s}=300\gev$ to approximate
the final $D_{c\rightarrow\jpsi}^{NLO}(z)$. It is clearly shown in Fig.~\ref{fig:frag} that $\jpsi$ from
the fragmentation is of larger momentum fraction at the NLO than that at the LO.

On the other hand, $D_{c\rightarrow\jpsi}^{NLO}(z)$ can be expressed as
\bea
D_{c\rightarrow\jpsi}^{NLO}(z)&=&D_{c\rightarrow\jpsi}^{LO}(z)\\
&\times&\left\{1+\dfrac{\a_s(\mu_R)}{\pi}\left[a(z)+\beta_0\ln \dfrac{\mu_R}{2m_c} \right]\right\}, \NO
\eea
where $\beta_0$ is the one-loop coefficient of QCD beta function. A 9th-polynomial fitting gives
$a(z)=\sum\limits_{i=0}^9c_iz^i$ with ~~~~~~
$c_0=-4.14253418603\times10^1,c_1=+1.08551074161\times10^2,$
$c_2=+1.62953354180\times10^3,c_3=-1.77039359042\times10^4,$
$c_4=+7.69974017791\times10^4,c_5=-1.82373533780\times10^5,$
$c_6=+2.52868747876\times10^5,c_7=-2.03305581610\times10^5,$
$c_8=+8.66677519749\times10^4,c_9=-1.47927072151\times10^4.$
This is the NLO fragmentation function of charm into $\jpsi$ at the initial
factorization scale $\mu_F=3m_c$. The fragmentation function at other factorization
scale can be obtained with this initial one by solving the Altarelli-Parisi evolution equation \cite{Altarelli:1977zs}.
In Fig.~\ref{fig:frag}b the evolutions of the LO and NLO fragmentation functions are presented.

It is easy to do applications with this new fragmentation function. Firstly we study
the decay of $Z^0$ into $J/\psi$ in $Z^0\rightarrow \jpsi+c\bar{c}$. With the additional parameter
$\a=1/128$ , the decay width at LO is
\bea
\Gamma^{LO}_{\jpsi+X}
=2\Gamma^{LO}_{c+X}\int dzD^{LO}_{c\rightarrow J/\psi}(z)=129 \kev. \NO
\eea
At NLO, there are two ways to calculate the decay width. One is described by Eq.~(\ref{eqn:NLO})
where the higher order term is neglected
\bea
&&\Gamma^{NLO}_{\jpsi + X}
=2\Gamma^{LO}_{c+X} \int dzD^{NLO}_{c\rightarrow \jpsi}(z) \NO\\
&&+2\int dE_c d z \dfrac{d\Gamma^{NLO*}_{c+X}}{d E_c} D^{LO}_{c\rightarrow \jpsi}\left(z\right)
=136\kev.\NO
\eea
The other one is to include the higher order term as
\bea
\Gamma^{NLO+}_{\jpsi+X}
=2 \int d E_c dz \dfrac{d\Gamma^{NLO}_{c+X}}{d E_c} D^{NLO}_{c\rightarrow \jpsi}(z)
=141\kev. \NO
\eea
Both the LO and NLO results are consistent with the fully NLO QCD calculation \cite{Li:2010xu},
which gives 120 KeV at LO and 136 KeV at NLO with same parameters. The differences come from the fact that
the limitation is not so well as the mass of $Z^0$ is not large enough to be treated as infinity.
Secondly we apply the fragmentation function to $b$ quark case by substituting
\bea
m_c \leftrightarrow m_b,~~n_f=4 \leftrightarrow n_f=5,~~R_s^\jpsi(0) \leftrightarrow R_s^\Upsilon(0). \NO
\eea
For the top quark decay, $t\rightarrow \Upsilon+W^+ +b$, we have
\bea
\Gamma^{LO}_{t\rightarrow \Upsilon+X}
=\Gamma^{LO}_{t\rightarrow b+X} \int d z D^{LO}_{b\rightarrow \Upsilon} (z)=30.9 \kev.
\label{eqn:twb0}
\eea
And the two corresponding NLO results are
\bea
&&\Gamma^{NLO}_{t\rightarrow \Upsilon+X}
=\Gamma^{LO}_{t\rightarrow b+X} \int dzD^{NLO}_{b\rightarrow \Upsilon}(z) \NO\\
&&+\int d E_b dz \dfrac{d\Gamma^{NLO*}_{t\rightarrow b+X}}{d E_b} D^{LO}_{b\rightarrow \Upsilon}\left(z\right)
=40.0\kev, \NO\\
&&\Gamma^{NLO+}_{t\rightarrow \Upsilon+X}=
\int dE_b dz \dfrac{d\Gamma^{NLO*}_{t\rightarrow b+X}}{d E_b} D^{NLO}_{b\rightarrow \Upsilon}\left(z\right)
=39.7\kev. \NO
\eea
Here we choose the same parameters as those used in Ref.~\cite{Sun:2010rw}. One should notice that the LO
wave function at the origin is used in Eq.~(\ref{eqn:twb0}), while in other two results the NLO one is used.
The corresponding LO and NLO results given by Ref.~\cite{Sun:2010rw} are 26.8 and 52.3 KeV.

The most important application is to study the production of $J/\psi+c\bar{c}+X$ at the LHC.
Because the fragmentation function $D_{g\to J/\psi}$ is suppressed comparing with $D_{c(\bar{c})\to J/\psi}$,
we only consider the contribution from the charm quark fragmentation.
The cross section is
\bea
&&\sigma[pp\to \jpsi c\bar{c}+X]\NO\\[3mm]
&=&\sum_{i,j=g,q,\bar{q}}\int dx_1dx_2dzf_{i/p}(x_1,\mu_f)f_{j/p}(x_2,\mu_f)\\[3mm]
&\times&d\hat{\sigma}[ij \to c\bar{c} +X,\mu_f,\mu_r,\mu_F]D_{c(\bar{c}) \to \jpsi}\left(z,\mu_F\right),\NO
\eea
%\begin{widetext}
%\bea
%\sigma[pp\to \jpsi c\bar{c}+X]
%=\sum_{i,j=g,q,\bar{q}}\int dx_1dx_2dzf_{i/p}(x_1,\mu_f)f_{j/p}(x_2,\mu_f) d\hat{\sigma}[ij \to c\bar{c} +X,\mu_f,\mu_r,\mu_F]D_{c(\bar{c}) \to \jpsi}\left(z,\mu_F\right),
%\eea
%\end{widetext}
where $f(x,\mu_f)$ is the parton distribution function, $\mu_f$, $\mu_r$ and $\mu_F$ are the factorization,
renormalization and fragmentation scales respectively. And $\hat{\sigma}$ represents the cross sections of partonic process.
The LO and NLO fragmentation functions are used to calculate the final LO and NLO results respectively. In the calculation,
there are two subprocesses $gg(q\bar{q})\to c\bar{c}$ at the LO,
and three real corrections $gg\to c\bar{c}+g$, $gq(\bar{q})\to c\bar{c}+q(\bar{q})$, $q\bar{q}\to c\bar{c}+g$ and
the virtual corrections to $gg(q\bar{q})\to c\bar{c}$ at the NLO.
The default choice of charm quark mass is $m_c=1.5$GeV and the three scales are set as $\mu_f=\mu_r=\mu_F=\mu$ with the
default choice $\mu=\mu_0=\sqrt{P_{t}^{c(\bar{c})}+m_c^2}$.
We choose  $m_c=1.4\gev,~1.6\gev$ and $\mu=\mu_0/2,~2\mu_0$ for uncertainty estimation.
The Cteq6L1 and Cteq6M~\cite{Pumplin:2002vw} are used in the LO
and NLO calculations respectively, with the corresponding $\alpha_s$ running formula being used.
The fragmentation function are evolved from $3m_c$ to $\mu_F$ by solving the Altarelli-Parisi equation numerically.
The wave function at the origin of $J/\psi$ is extracted from its leptonic decay as in reference \cite{Li:2010xu} at the NLO level and
the rapidity cut for $J/\psi$ is $|y_{J/\psi}|<2.4$.

\begin{figure}
\begin{center}
\includegraphics[scale=0.21]{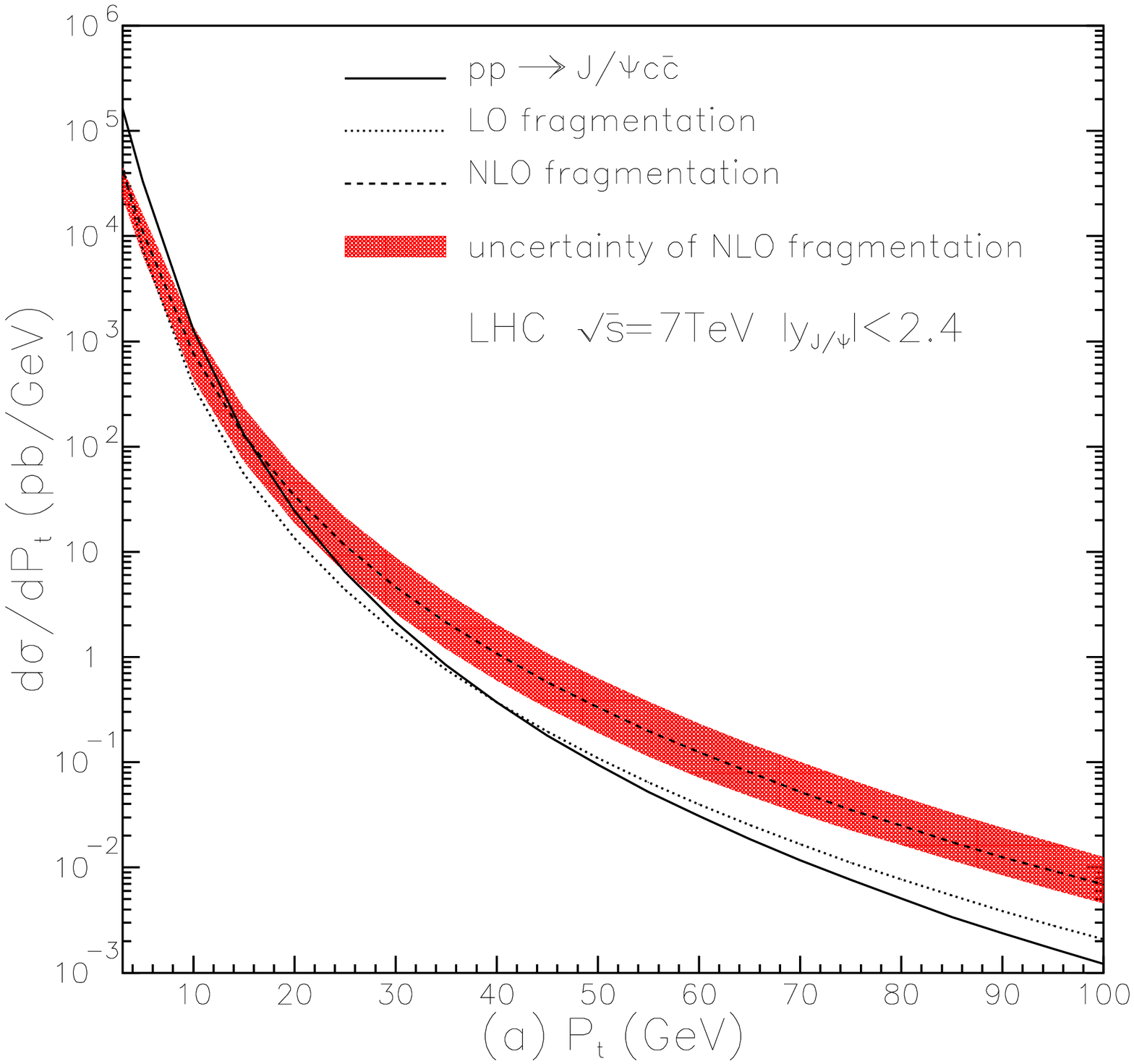}
~~\includegraphics[scale=0.21]{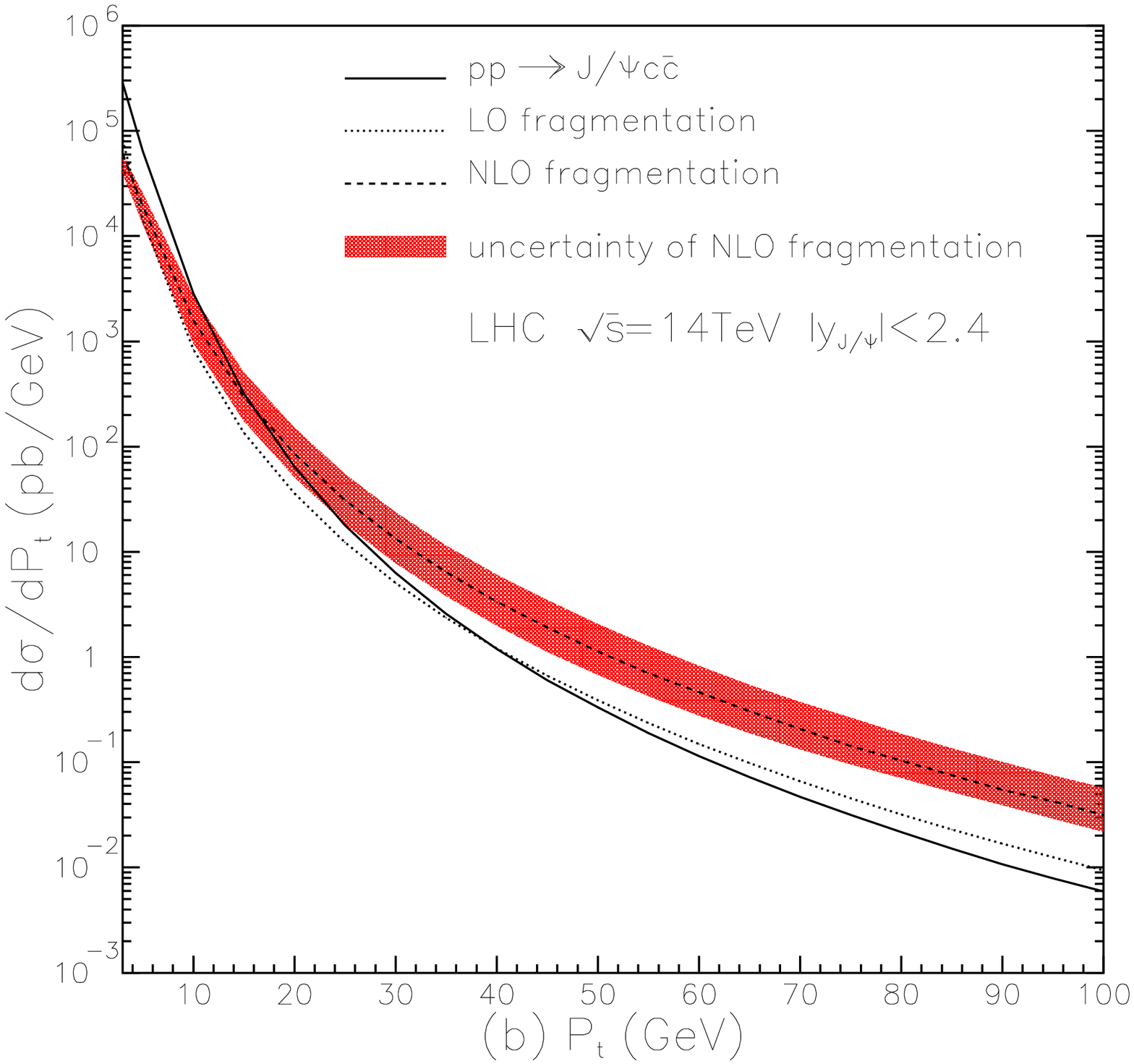}
\caption{\label{fig:pt}The theoretical prediction on $p_t$ distribution  of $J/\psi$ production
associated with a charm c ($\bar{c}$) jet at the LHC.}
\end{center}
\end{figure}

The predictions for transverse momentum distributions of $J/\psi$ are shown
in Fig.~\ref{fig:pt} with $\sqrt{s}=7$ TeV and $\sqrt{s}=14$ TeV at the LHC. The complete calculation of process
$pp\to J/\psi +c\bar{c}+X$ at the LO are presented for comparison, which was studied in reference
\cite{Artoisenet:2007xi}%{,Hagiwara:2007bq}
. It will be dominated over as $p_t\ge40\gev$ by the LO fragmentation result
and as $p_t\ge 16\gev$ by the NLO fragmentation one, and is about 5 times smaller than the NLO fragmentation one
in large $p_t$ region.  It is believed that the fragmentation mechanism should give a better
description on the $J/\psi$ hadroproduction associated with a charm c ($\bar{c}$) jet at large $p_t$ region
than the complete calculation result.

\label{summary}
We extracted the fragmentation function for charm into $\jpsi$ at the NLO in $\a_s$, $D_{c\rightarrow\jpsi}^{NLO}(z,3 m_c)$,
from $\jpsi$ production at $e^+e^-$ annihilation.  The fragmentation function for bottom into $\Upsilon$,
$D_{b\rightarrow\Upsilon}^{NLO}(z,3 m_b)$, is also obtained.
The Altarelli-Parisi evolution of the fragmentation function is performed to obtain its value at other fragmentation scale.
We have applied them to $\jpsi$ production in $Z^0$ decay, and $\Upsilon$ production in top quark decay.
The most important application is to predict the transverse momentum distribution of $J/\psi$ production associated with a charm c
(or $\bar{c}$) jet at the LHC.
We find that the distribution is enhanced by a factor of 2.0$\sim$3.3 at the NLO as $p_t$ increased from 10 GeV to 100 GeV
and it is measurable at the LHC with charm tagger.  In the measurement, it need to identify a $J/\psi$ and a charm jet
without the $J/\psi$ in the jet and there will be 20 events to be found if we optimistically assume a $60\%$ charm tagging efficiency,
both $e$ and $\mu$ decay channel of $J/\psi$ being detected and $10fb^{-1}$ of the integrated luminosity at a 14 TeV machine.
Of course, this measurement is a very big challenge to the experimental technique on charm tagger. However,
the measurement at the LHC will supply a first chance to directly
test the QCD fragmentation mechanism on heavy quarkonium production where the fragmentation function is calculable
in perturbative QCD.

We thank Y. Jia, Y.Q. Chen and Q. S. Yan for helpful comments and discussions.
This work was supported by the National Natural Science Foundation of China (No.~10979056, 10935012 and 11005137), and the China Postdoctoral Science Foundation (No.~20090460535 and 20090460525).
%\appendix
%\section{Appendixes}
%\newpage %Just because of unusual number of tables stacked at end
\bibliography{paper}% Produces the bibliography via BibTeX.
\end{document}